\documentclass[letterpaper,english,aps,prb,twocolumn,floatfix,showpacs,amsfonts,amssymb,superscriptaddress]{revtex4}

\usepackage[T1]{fontenc}
\usepackage[latin1]{inputenc}
\usepackage{graphics}
\usepackage{amssymb}
\usepackage{overpic}

\newcommand{\beq}{\begin{equation}}
\newcommand{\eeq}{\end{equation}}
\newcommand{\bea}{\begin{eqnarray}}
\newcommand{\eea}{\end{eqnarray}}

\newcommand{\rmi}{{\rm i}}

\newcommand{\lco}{La$_{2}$CuO$_{4}$}
\newcommand{\lasco}{La$_{2-x}$Sr$_{x}$CuO$_{4}$}

\begin{document}

\title{Strong correlations and the anisotropy of acceptor states in insulating $\mbox{\lasco}$}


\author{M.~B.~Silva~Neto}

\affiliation{Instituto de F\' isica, Universidade Federal do
Rio de Janeiro, Caixa Postal 68528, Brasil}

\begin{abstract}

We use the Green's function formalism to discuss the role of strong 
correlations to the spatial structure of acceptor states doped into a 
two-dimensional Mott-Hubbard antiferromagnetic insulator. When 
the scattering between doped carriers, at the nesting wave vector 
${\bf Q}=(\pi,\pi)$, is strong enough to produce a momentum 
dependent scattering rate, $\Gamma_{\bf k}$, the corresponding 
acceptor states become spatially anisotropic. As an example, we 
calculate the spatial structure of an acceptor state bound to an 
attractive two-dimensional Dirac delta potential, for a simple form
of $\Gamma_{\bf k}$. We then discuss the role of such spatial anisotropy 
for the understanding of an apparent discrepancy between low 
temperature transport data and photoemission spectra 
in lightly doped {\lasco}.

\end{abstract}


\maketitle

The quantitative description of hopping conductance
in extrinsic semiconductors relies on what is called the
effective mass approximation \cite{Cardona}. Once the band structure
of the host material is known, the effective masses on
top of the valence band, for acceptors, or bottom of the
conduction band, for donors, can be calculated, and a 
Schr\"odinger equation for the trap potential, generated 
by the impurity ions on the doped carriers, is then written 
down \cite{ES-Book}. The bound states associated with 
the negative energy solutions to this equation correspond 
to localized states and, usually, the envelope function 
at the lowest energy state is spatially isotropic, symmetric,
parity even, just like the Hydrogen atom 1s state. The Bohr 
radius of the bound state (or inverse effective mass) 
controls the exponential decay of the envelope wave 
function, the overlap between two spatially separated 
impurity states, and directly affects the hopping 
conductance.

Undoped cuprates, such as {\lco}, are far from being 
classified as extrinsic semiconductors. These insulating 
materials actually belong to the class of strongly correlated 
electron systems, and exhibit long ranged antiferromagnetic 
order in the ground state. Nevertheless, it has been found 
that, when doped with a small amount of carriers, {\lasco}, 
for example, exhibits metallic behavior at high temperature 
(albeit anomalous) already for $1\%$ of carriers \cite{Exp-Ando}. 
Furthermore, {\lasco} also exhibits hopping conductance at 
low temperature, in a very similar fashion as described above. 
In particular, at $x=3\%$ and $x=4\%$, both AC \cite{Dumm-Padilla} 
and DC  \cite{DC-Resistivity} data are consistent with hopping 
transport and point towards the existence of localized acceptor 
states. 

Contrary to the above tendency, recent angular resolved 
photoemission spectroscopy (ARPES) experiments in 
{\lasco} revealed a certain degree of ``metalicity'' at the 
same doping regime discussed above \cite{Yoshida}, 
and have, as a result, been interpreted as evidence for 
delocalization of the charge carriers. The observation of 
``quasiparticle'' peaks in the form of small Fermi arcs for 
$3\%$ doped {\lasco} at $T=10K$ \cite{Yoshida}, which 
according to transport should be insulating, resembles 
very much the full pocket structure observed in other 
cuprates at higher doping \cite{Full-Pockets}, where the 
sample is (strangely) metallic. The conclusion was that, 
since ARPES shows ``peaks'' in reciprocal (momentum) 
space, the associated wave function should be extended 
in real (coordinate) space, and not localized as suggested 
by hopping transport. 

As an attempt to address this apparent discrepancy, A. S. 
Alexandrov and K. Reynolds \cite{Alexandrov} suggested 
that such Fermi arcs observed with ARPES in insulating
{\lasco} could arise naturally from matrix elements when 
the ejected electron leaves behind a hole at {\it valence band 
tails}, formed by the hybridization of valence band and impurity 
states \cite{ES-Book}. The reasoning is actually relatively simple.
 While long-lived quasiparticles in a Fermi 
liquid have very well defined momentum states and thus give 
rise to sharp peaks in the ARPES spectrum, localized wave 
functions are written as a linear superposition of many different 
wave vectors, and, as a consequence, do not possess a well 
defined momentum state. For this reason, no peaks are to be 
expected in the ARPES spectrum corresponding to a bound
state. For a sufficiently large envelope wave function, however,
the ARPES matrix elements in reciprocal space will depend 
on which of the linearly combined wave vectors weigh a 
heavier contribution to the spectral function. For the case 
of cuprates, for example, it is argued, in Ref.\ \onlinecite{Alexandrov},
that the wave vectors close to $(\pi/2,\pi/2)$  contribute a 
much higher spectral weight, which then falls rapidly away 
from this point \cite{Alexandrov}. For this reason, bound 
electrons ejected from band tails are able to produce 
Fermi arcs in the ARPES spectrum. 

A necessary condition for the Alexandrov and Reynolds idea 
to work is that the localized state should be anisotropic in real 
space \cite{Alexandrov}, being more elongated along the 
direction perpendicular to the Brillouin zone (B.Z.) faces, 
and being shorter along the direction parallel to it \cite{Alexandrov}. 
In this work we extend the traditional effective mass approximation 
to include strong correlations, and we calculate the full spatial 
structure of acceptor states using a phenomenological model to 
mimic light doping in {\lasco}. We argue that due to the 
presence of a nesting AF wave vector $Q=(\pi,\pi)$ (in units of 
the lattice spacing $a=1$) the localized states become squeezed 
exactly along the B.Z. faces and elongated along the so called 
nodal directions (along the direction perpendicular to the B.Z. 
faces), providing the missing ingredient to complete the 
Alexandrov and Reynolds analysis.

We begin by reviewing the effective mass approximation
for the calculation of acceptor/donor states in extrinsic 
semiconductors \cite{ES-Book}. The effective mass tensor, 
$m^*_{\alpha\beta}$, is defined as 
$(m^*_{\alpha\beta})^{-1}=\partial^2\varepsilon({\bf k})/
\partial k_{\alpha}k_{\beta}$, where $\varepsilon({\bf k})$ 
is the quasiparticle dispersion close to the bottom of the 
conduction band (for donors) or top of the valence band 
(for acceptors) located at the $i$-th pocket. In case the 
effective mass tensor is diagonal, but with different 
components, $m^*_\parallel$ and $m^*_\perp$, the 
time-independent Schr\"odinger equation for the localized 
wave function $\psi^i({\bf r})$ reads
\beq
\left[-\left(\frac{\hbar^2\nabla_\parallel^2}{2m_\parallel^*}
+\frac{\hbar^2\nabla_\perp^2}{2m_\perp^*}\right)+V({\bf r})+
U({\bf r})\right]\psi^i({\bf r})=E\; \psi^i({\bf r}),
\label{Schroedinger}
\eeq
where $\nabla_\perp=\partial/\partial z$,
$\nabla_\parallel=(\partial/\partial x,\partial/\partial y)$,
$V({\bf r})$ is the periodic potential provided by the
lattice, and $U({\bf r})$ is the trap potential felt by the 
doped carriers and provided by an impurity at the 
origin of the coordinate system. For large enough
bound states, when the localization length is of
the order of several lattice spacings, one can 
write
\beq
\psi^i({\bf r})=F^i({\bf r})\phi^i({\bf k}^i,{\bf r}),
\eeq
where $F^i({\bf r})\sim e^{-r/\xi^i}$ is usually an 
exponentially decaying envelope wave function with
localization length $\xi^i$ at the $i$-th pocket and, 
according to Bloch's theorem,
$\phi^i({\bf k}^{i},{\bf r})=e^{\rmi{\bf k}^i \cdot{\bf r}} u_{{\bf
    k}^i}({\bf r})$,
where $u_{{\bf k}^i}({\bf r})$ is a periodic 
function with minimum at ${\bf k}^{i}$. 
Following Kohn 
and Luttinger (KL)  \cite{Kohn-Luttinger}, each one
of the $\mu=1\dots N$ degenerate (from different
pockets) localized states are generally written as
\beq
\Psi_{\mu}=\sum_{i=1}^N\alpha_\mu^i \psi^{i}({\bf r})=
\sum_{i=1}^N\alpha_\mu^i F^i({\bf r})\phi^i({\bf k}^i,{\bf r}),
\eeq
where $\alpha_\mu^i$ are $N$-dimensional vectors
determined by point group symmetry carrying the
information about the symmetries of the wave functions.
We define $\gamma=m^*_\perp/m^*_\parallel$,
and $F^i({\bf r})\sim e^{-r/\xi^i}$ can be
classified as isotropic $3$-D, $\gamma=1$, in which
case $r=\sqrt{x^2+y^2+z^2}$, or extremely adiabatic 
$2$-D, $\gamma\rightarrow 0$, in which case 
$r=\sqrt{x^2+y^2}$.

%
\begin{figure}[t]
\includegraphics[scale=0.36]{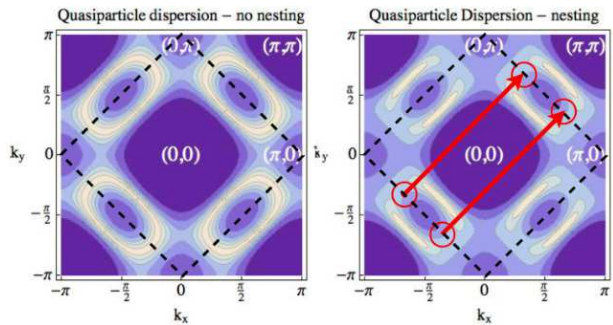}
\caption{(Color online) Evolution of the quasiparticle
spectrum: a) (left) fully coherent hole-like Fermi pockets
 \cite{Full-Pockets}, located at ${\bf k}_0=(\pm\pi/2,\pm\pi/2)$ 
in the magnetic B. Z.; b) (right) additional scattering at the 
nesting antiferromagnetic wave vector \cite{Hot-Spots}, 
${\bf Q}=(\pi,\pi)$, and the {\it hot spots} along the magnetic 
B.Z. faces (effect of a ${\bf k}$-dependent scattering rate 
${\Gamma}_{\bf k}$).}
\label{Fig-Evolution-Pockets}
\end{figure}
%

It has long been acknowledged that, to a very good extent, 
the physics of lightly doped cuprates is well captured by 
the $t-t^\prime-t^{\prime\prime}-J$ model \cite{Review}, 
which includes not only nearest neighbor hopping, $t$, 
but also second neighbor hoppings, $t^\prime$ and 
$t^{\prime\prime}$, for holes doped into a two-dimensional 
antiferromagnet with superexchange $J$. At a very small 
concentration, the Fermi surface for the added holes 
correspond to four, nearly two-dimensional hole pockets 
centered at the wave vectors ${\bf k}_0=(\pm\pi/2,\pm\pi/2)$ 
in the magnetic B.Z.,  as can be seen in 
Fig.\ \ref{Fig-Evolution-Pockets}a). This is consistent 
with band structure calculations, as well as ARPES 
data  \cite{Full-Pockets}, and the two-dimensionality 
of the Fermi surface results from the huge effective 
along the direction perpendicular to the CuO$_{2}$ 
planes (or parallel to the $c$-axis), 
$m^*_\perp \gg m^*_\parallel$, where 
$m^*_\parallel$ is the planar effective mass. The 
above large effective mass anisotropy drives the 
system into the extreme adiabatic limit discussed 
above, $\gamma\rightarrow 0$. Self consistent Born 
approximation for the $t-t^\prime-t^{\prime\prime}-J$ 
model has also determined that, at the very top of 
the valence band, the pockets are nearly circular, 
with equal effective masses along the magnetic B.Z. 
faces and diagonals, or orthorhombic $(a,b)$ axis,
and thus we can set $m^*_a\simeq m^*_b=m^*_\parallel$. 
In this case
\beq
F({\bf r})\sim e^{-r/\xi},
\eeq
where $r=\sqrt{x^2+y^2}$ and $\xi\sim 1/m^*_\parallel$. 
Such large effective mass anisotropy, 
$m^*_\perp\gg m^*_\parallel$ is consistent with 
the large resistivity anisotropy in these materials 
\cite{DC-Resistivity}, $\rho_c/\rho_{ab} \sim 10^3$. 

However, a 2D isotropic wave function is not 
consistent with neither the ARPES response for 
very lightly doped {\lasco} nor with the AC and DC 
transport measurements at low temperature, in the 
hopping regime, where the knowledge of the precise 
shape of the localized state turns out to be crucial.
These experiments have revealed that, besides the 
perpendicular anisotropy discussed above, 
$m^*_\perp/m^*_\parallel\gg 1$, an additional 
in-plane anisotropy is necessary to account for the 
different infrared absorption spectra observed along 
the two orthorhombic directions in {\lasco} 
\cite{Dumm-Padilla}. In what follows, we shall argue 
that, even more important than the anisotropy caused 
by spiral correlations considered by V. Kotov and O.
Sushkov in Ref.\ \onlinecite{Kotov}, is the anisotropy 
coming from strong correlations, in the form of a 
momentum dependent scattering rate 
\cite{Hot-Spots,Anisotropic-Scattering}. This will
force us to extend the usual effective mass
approximation, used in Eq. (\ref{Schroedinger}), 
in order to incorporate such nontrivial effect.
 
On general grounds, we shall assume that the original 
quasiparticle dispersion of the noninteracting, $J=0$, 
problem, the $t-t^\prime-t^{\prime\prime}$ model, have 
the pocket like structure at ${\bf k}_0=(\pm\pi/2,\pm\pi/2)$ 
shown in Fig. \ref{Fig-Evolution-Pockets}a). For $J\neq 0$, in turn,
short wavelength antiferromagnetic fluctuations, at low
temperature and near half-filling, cause the enhancement 
of the quasiparticle scattering at the ordering wave vector 
\cite{Hot-Spots}, see Fig. \ref{Fig-Evolution-Pockets}b). 
The presence of such {\it hot spots} produces, in turn, 
a momentum dependent scattering rate (or inverse 
quasiparticle lifetime) \cite{Anisotropic-Scattering}, 
which provides us with an important source of anisotropy. 

For simplicity, let us consider the problem
of a hole under the influence of an attractive 2D-delta 
potential\cite{Negative-MR}. According to the usual effective mass approximation, 
the envelope wave function $F({\bf r})$ can be obtained from
\beq
\left[-\frac{\hbar^2\nabla^2}{2m^*_\parallel}+U({\bf r})\right]F({\bf r})=E\; F({\bf r}),
\label{Main-Equation}
\eeq
where 
\beq
U({\bf r})=-g\frac{\delta({\bf r}-{\bf r}_0)}{r}
\eeq
is an attractive $\delta$-potential that traps, with strength
$g$, a hole around an impurity ion at the origin. 
We shall be interested in the solutions to this equation
for ${\bf r}$ larger than the cutoff ${\bf r}_0$. Using
the Fourier transform
\beq
F({\bf r})=\int\frac{d^2{\bf k}}{(2\pi)^d}\;e^{-\rmi{\bf k}\cdot{\bf
    r}}\; f({\bf k}),
\eeq
the equation becomes
\beq
\int\frac{d^2{\bf k}}{(2\pi)^d}
\left[\frac{\hbar^2{\bf k}^2}{2m^*_\parallel}-E\right]
e^{-\rmi{\bf k}\cdot{\bf r}}\; f({\bf k})=g
\frac{\delta({\bf r}-{\bf r}_0)}{r}F({\bf r}).
\eeq
We now multiply the whole equation by $(2\pi)^2\; e^{\rmi{\bf q}\cdot{\bf r}}$, 
and integrate over ${\bf r}$ to obtain
\beq
f({\bf q})=
\frac{4\pi^2g_0\; F(r_0)}{\varepsilon({\bf q})+\varepsilon_B}=
g_0\;\frac{f(r_0)}{\varepsilon({\bf q})+\varepsilon_B},
\label{First-BS-Eq}
\eeq
where we defined $g_0=g/r_0$, $f(r_0)=4\pi^2 F(r_0)$, and
$\varepsilon({\bf q})=\hbar^2{\bf q}^2/2m^*_\parallel$, is the dispersion 
in the effective mass approximation. Here we have used, as 
the negative energy (bound state) solution, 
$-E=\varepsilon_B=\hbar^2{\bf \kappa}^2/2m^*_\parallel$, with 
$\kappa=1/\xi$ playing the role of inverse localization length 
$\xi$. We recognize Eq. (\ref{First-BS-Eq}) as a particular 
case of the more generic Bethe-Salpeter equation for a potential
$U({\bf r})$
\beq
f({\bf q})=-
\frac{4\pi^2}{\varepsilon({\bf q})+\varepsilon_B}
\int\frac{d^2{\bf k}^\prime}{(2\pi)^2}
\tilde{U}({\bf k}^\prime-{\bf q})\; f({\bf k}^\prime),
\eeq
with $\tilde{U}({\bf k})$ being, as usual, the Fourier transform
of $U({\bf r})$.

The result presented in (\ref{First-BS-Eq}) is the exact
solution to the equation (\ref{Main-Equation}) in which
all wave vector are treated on equal footing. In particular,
the term $\hbar^2\nabla^2/2m^*_\parallel$ is isotropic and results
from the nearly perfect circular shape of the hole pockets
in lightly doped {\lasco}.  For this reason, $f({\bf q})$ depends 
only on ${\bf q}^2$ and the resulting envelope function
$F({\bf r})$ is also isotropic. In order to introduce the 
momentum dependence of the scattering rate \cite{Hot-Spots}, 
$\Gamma_k$, a quantity that is related to the quasiparticles 
spectral function and which does not appear in (\ref{Main-Equation}), 
we will make use of the Green's functions formalism, and its 
spectral representation. We perform a Stieltjes transform to 
write
\beq
G({\bf q},\omega=0)=\frac{1}{\varepsilon({\bf q})+\varepsilon_B}=
\int_{-\infty}^{\infty} dE\;\frac{A({\bf q},E)}{E+\varepsilon_B},
\eeq
where the spectral function $A({\bf q},E)$ is, for such strongly
correlated electron system, given by
\beq
A({\bf q},E)=\frac{{\cal Z}_{\bf q}}{\pi}
\frac{\Gamma_{\bf q}}{(E-\varepsilon({\bf q}))^2+\Gamma_{\bf q}^2}
+A_{inc}({\bf q},E),
\eeq
where $Z_{\bf q}$ is the quasiparticle weight, $\Gamma_{\bf q}$ is 
the momentum dependent scattering rate, and $A_{inc}$ is 
the incoherent (or multiparticle) part of the spectrum. For a weakly
interacting system, a Fermi liquid, where $\Gamma_{\bf q}=
\Gamma_0\rightarrow 0$, one ends up with
$A({\bf q},E)=Z_{\bf q}\delta(E-\varepsilon({\bf q}))+A_{inc}({\bf q},E)$.

%
\begin{figure}[t]
\includegraphics[scale=0.34]{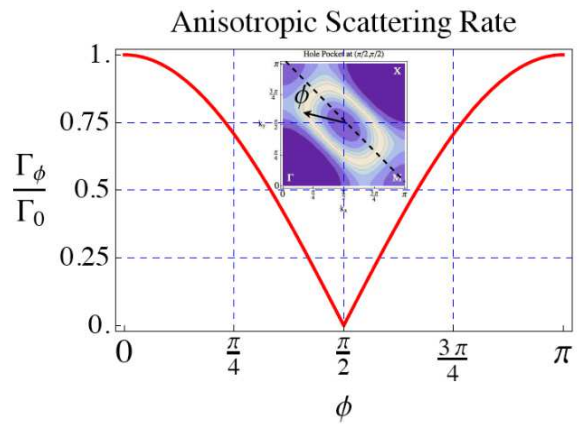}
\caption{(Color online) Anisotropic scattering rate $\Gamma_\phi$
(red solid line) as a function $\phi$. Inset: $\phi$ is the 
angle between the wave vector ${\bf q}$ (black solid arrow) and 
the magnetic B.Z. face (black dashed line). At the hot spots, 
$\phi=0,\pi$, the scattering is maximal, while at the nodal 
direction, $\phi=\pi/2$, it is minimal \cite{Anisotropic-Scattering}.}
\label{Fig-Scattering-Rate}
\end{figure}
%
Since the nesting is strongest for the points where the pockets 
intersects the B.Z. faces, we can, without loss of generality, consider 
the phenomenological formula
\beq
\Gamma_{\bf q}=\Gamma_0\left|\cos{(\phi)}\right|,
\eeq
where $\phi$ is the angle between the wave vector ${\bf q}$,
centered at a given pocket, and the B.Z. face. This simple form 
of the scattering rate produces larger widths for the spectral lines 
along the B.Z. faces, the hot spots at $\phi=0,\pi$, very small 
widths along the nodal directions, $\phi=\pi/2$ (see 
Fig.\ \ref{Fig-Scattering-Rate}), and is consistent with the
results from Ref.\ \onlinecite{Anisotropic-Scattering}. 

For calculating bound states, it is enough to keep only the coherent part of the spectrum, defined as 
$A_{coh}({\bf q},E)=A({\bf q},E)-A_{inc}({\bf q},E)$, and thus we write
\beq
F({\bf r})=
g_0 f(0)
\int\frac{d^2{\bf q}}{(2\pi)^2}
\int dE\frac{e^{-\rmi{\bf q}\cdot{\bf r}}}{E+\varepsilon_B}A_{coh}({\bf q},E).
\eeq
We now make the following approximations. We use explicitly that 
$Z_{\bf q}=Z_0$, that is, the quasiparticle weight is weakly momentum 
dependent at low doping, and we use explicitly the angular dependence 
of the scattering rate, $\Gamma_{\bf q}=\Gamma_{\phi}$. We evaluate this 
integral by closing a contour in the upper half plane and use the residue 
theorem
\beq
F({\bf r})=g_0 f(r_0){\cal Z}_0
\int\frac{d^2{\bf q}}{(2\pi)^2}\; e^{-\rmi{\bf q}\cdot{\bf r}}
\frac{\varepsilon({\bf q})+\varepsilon_B}{(\varepsilon({\bf q})+\varepsilon_B)^2+\Gamma_\phi^2},
\eeq
where we have used explicitly that ${\cal Z}_{\bf q}={\cal Z}_{0}$ and
$\Gamma_{\bf q}=\Gamma_\phi$. Notice that in the limit
where $\Gamma_{\phi}\rightarrow 0$ and ${\cal Z}_{0}\rightarrow 1$
the above equation reduces to the result obtained in (\ref{First-BS-Eq}).

\begin{figure}[t]
\includegraphics[scale=0.34]{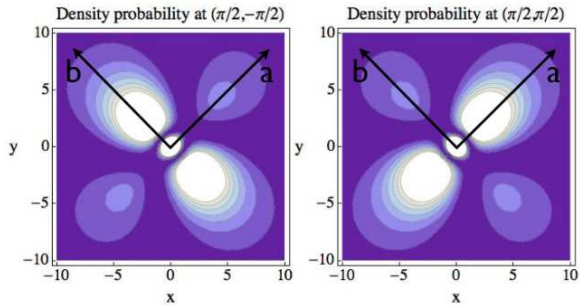}
\caption{(Color online) Density plot of the amplitude
probability for the envelope wave function $F({\bf r})$, 
at the two nonequivalent pockets: a) (left) $(\pi/2,-\pi/2)$;
and b) (right) $(\pi/2,\pi/2)$. The axis $a$ and $b$
are the orthorhombic axis.}
\label{Fig-Density-Plot}
\end{figure}

Before performing the angular integration, we must recall that, while $\phi$
is the angle between ${\bf q}$ and the B.Z. face, the angle $\theta$ that 
appears in ${\bf q}\cdot{\bf r}=qr\cos\theta$, is the angle between ${\bf q}$ 
and ${\bf r}$. So, if we define as $\varphi$ the angle between ${\bf r}$ and 
the B.Z. face at a given pocket, we end up with $\theta=\phi-\varphi$. The complete solution 
to the bound state problem can be finally presented.
We approximate $\Gamma_\phi\ll \varepsilon_B$, and write
$F_\varphi({\bf r})=F_0({\bf r})+\delta F_\varphi({\bf r})$,
where
\beq
F_0(\kappa,{\bf r})=g_0 f(r_0){\cal Z}_0\frac{m^*_\parallel}{\pi\hbar^2} K_0(\kappa r),
\eeq
is the unperturbed ($\Gamma_0= 0$) isotropic 
($\varphi$-independent) contribution to the wave function, while 
the perturbation (and source of anisotropy in $\varphi$) becomes
\begin{widetext}
\beq
\delta F_\varphi(\kappa,{\bf r})=g f(r_0){\cal Z}_0\Gamma_0^2\;
\frac{2 (m^*_\parallel)^2}{\kappa\hbar^4}
\left\{
\frac{2 \left(2-\kappa^2 r^2 K_0(\kappa r)\right) 
\cos{(2 \varphi)}-\kappa r K_1(\kappa r)\left(\kappa^2 r^2+
\left(4+\kappa^2 r^2\right) 
\cos{(2 \varphi)}\right)}{2 \kappa^3 r^2}
\right\},
\eeq
\end{widetext}
where $K_0$ and $K_1$ are modified Bessel functions of the
second kind and, from its definition, $\varphi$ is given by
$\varphi=\arctan[(x\mp y)/(x\pm y)]$, for the pockets centered
at $(\pi/2,\pm \pi/2)$, see Fig.\ \ref{Fig-Density-Plot}. 
We see that, as required by Alexandrov and Reynolds \cite{Alexandrov}, 
the spatial structure of the acceptor state is such that the envelope 
wave function is elongated exactly along the nodal directions, or
perpendicular to the B.Z. faces, and squeezed along the directions
parallel to the B.Z. faces. 

%
%

We have extended the traditional effective mass approximation\cite{ES-Book}
to include the effects from strong correlations, reflected in
the anisotropy of the quasiparticle scattering rate\cite{Hot-Spots}, 
$\Gamma_{\bf k}$ (or inverse quasiparticle lifetime). Using the 
Green's function method, and its spectral representation, we 
were able to calculate the full structure of an acceptor state 
trapped by a simple two-dimensional $\delta$-potential, used to 
mimic the problem of lightly doping a Mott-Hubbard antiferromagnetic 
insulator. Our results provide the missing ingredient to the 
Alexandrov and Reynolds analysis\cite{Alexandrov}, which bridges 
the gap between low temperature transport\cite{Dumm-Padilla,DC-Resistivity} 
and ARPES\cite{Yoshida} data.

The author acknowledges discussions with A. Alexandrov,
G. Blumberg, R. Capaz, A. Chernyshev, B. Koiller, and O. Sushkov.

\end{document}